\documentclass[12pt]{iopart}
\begin{document}
\title{Relativistic hydrodynamics with strangeness production}
\author{L Turko}
\address{Institute of Theoretical Physics, University of
Wroclaw,\\ pl. Maksa Borna 9, 50-204 Wroclaw, Poland}\ead{turko@ift.uni.wroc.pl}
\begin{abstract}
The relativistic hydrodynamic approach is used to describe production of strangeness and/or heavy quarks in ultrarelativistic heavy ion reactions. Production processes are important ingredients of dissipative effects in the hadronic liquid. Beyond viscosity also chemo- and thermo-diffusion processes are considered. This also allows to specify chemical and thermal freeze-out conditions.
\end{abstract}
\submitto{\jpg}

\section{Introduction}
Results of RHIC experiments showed up that data of ultrarelativistic heavy ion collisions can be described using relativistic hydrodynamic of a thermalized perfect fluid \cite{rhic, qgp3}. Indications came mainly from a large elliptic flow effect. It appears, however, that dissipative effects become more and more important \cite{heinz} beyond mid-rapidity, most central, and the top RHIC energy regions. A more detailed analysis of initial conditions of color glass condensate models needs a finite viscosity to reduce the elliptic flow to match the data \cite{adil}. There are also some general arguments that at least shear viscosity of matter created in heavy ion collision must be nonzero \cite{policastro,kapusta}.

The entropy of the QGP phase at hadronization determines the observable particle multiplicity \cite{bjo83,Let93,Let94} so it is important to take into account contributions due to particle production processes. The entropy content is usually estimated assuming thermal momentum distributions and related equations of state. A more general formalism allowing for the consistent incorporation of dissipative and entropy producing mechanisms into the relativistic hydrodynamical framework was presented in \cite{ert}. That was the hydrodynamic model with dynamical chemical reactions and multi-fluid included. We are going to specify here this general scheme for the case of strangeness production in the quark-gluon plasma (QGP) case.

The entropy balance is treated here in the deconfined QGP phase. The primary contribution is from the initial decoherence processes which however cannot be calculated \emph{ab initio.} This contribution is estimated from 40 \cite{Geiger94} to 70 \% \cite{Let94:a} of the total  entropy produced. We consider the contributions due to the production of secondary partons. This is the main \emph{calculable} entropy source. The entropy originated from the hadronization processes is rather small.

The energy-momentum tensor for an imperfect fluid is obtained in the standard procedure \cite{LandauLif2}. The energy-momentum tensor is given as
\begin{equation}\label{em tensor 1}
T^{\mu \nu}\equiv (\epsilon +P)u^\mu u^\nu -Pg^{\mu \nu} +{\delta T^{\mu \nu}} \;,
\end{equation}
with the  term ${\delta T^{\mu \nu}}$ related to all dissipative processes in the fluid. $\epsilon$ and $P$ denote the local energy density and pressure, respectively.

The evolution of the closed system is described by equations
\begin{equation}
\partial _\mu T^{\mu \nu} = 0\,.
\end{equation}
These equations are supplemented by continuity equations
\begin{equation}\label{contin}
\partial _\mu\rho _i^\mu\equiv \partial _\mu  \rho_i u^\mu =0 \;\;,
\end{equation}
where $\rho_i$ denotes the particle number density in the local rest frame. The local rest frame is determined by the four-velocity condition $u^\mu=(1,0,0,0)$.

We require dissipative terms to vanish in the local rest frame,
\begin{equation}
T_{(0)}^{00}=\epsilon \;\;\longrightarrow\;\;\; \delta T_{(0)}^{00}=0 \;\;\longrightarrow\;\;\; u_\mu u_\nu\delta T^{\mu\nu}=0 \;\;.
\end{equation}
These conditions allow to determine the general structure of the dissipative  term $\delta T^{\mu\nu}$ \cite{ert}.
  \subsection{Diffusion, particle production, chemical reactions}
We generalize continuity equations \eref{contin} to incorporate diffusion and particle production or chemical reactions. \Eref{contin} is replaced now by
\begin{equation}
\partial _\mu \tilde\rho_i^\mu \equiv
\partial _\mu (\rho _i^\mu +\triangle^{\mu\nu}
\partial _\nu {\cal R}_i)={\cal J}_i
\;\;.
\end{equation}
with $\triangle_{\mu\nu}\equiv g_{\mu\nu}-u_\mu u_\nu,\ g_{\mu\nu}$ = diag(1,-1,-1,-1) the metric of the flat space-time.

$\mathcal{R}$ is the diffusion term which is determined consistently \cite{ert} with the Second Law of Thermodynamics. ${\cal J}_i$ are sources of particle production. They are determined by microscopic reaction cross sections. In the local rest frame one gets the gain and loss equation
\begin{equation}\label{gain loss}
{\cal J}_i= \sum _j[G_{i\leftarrow j}\rho _j-L_{j\leftarrow i}\rho _i] \;\;,
\end{equation}

Since the entropy of the QGP phase at hadronization determines the observable particle multiplicity, it is crucial to understand its production. The entropy production in a dissipative relativistic fluid is given by:
\begin{eqnarray}
T\partial _\mu s^\mu &=&-\sum_i\mu _i{\cal J}_i-T^{-1}(\partial _\mu T)u_\nu \delta T^{\mu\nu} +(\partial _\mu u_\nu )\delta T^{\mu\nu} \nonumber \\ &\;&+\sum_i(\mu _i+T{\cal L}_i)\partial _\mu\triangle ^{\mu\nu}\partial _\nu  {\cal R}_i +T\sum_i(\partial _\mu{\cal L}_i)\triangle ^{\mu\nu}\partial _\nu  {\cal R}_i \;\;,
\end{eqnarray}
where $T$ is the temperature, $u^2=1$, and we employed natural units $\hbar=c=k_B=1.$

The heat-flow four-vector is defined by
\begin{equation}\label{heat}
{\cal Q}_\mu \equiv \partial _\mu T-Tu^\nu\partial _\nu u_\mu \,,
\end{equation}
and the shear tensor is given by
\begin{equation}\label{shear}
{\cal W}_{\mu\nu}\equiv \partial _\mu u_\nu +\partial _\nu u_\mu -\frac{2}{3} g_{\mu\nu}\partial _\gamma u^\gamma \,.
\end{equation}
\section{Entropy Production}
The First Law of Thermodynamics imply:
\begin{equation}
0=T\partial _\mu\tilde{s}^\mu +\sum_i\mu _i({\cal J}_i-\partial _\mu \triangle ^{\mu\nu}\partial _\nu {\cal R}_i) +u_\nu\partial _\mu\delta T^{\mu\nu} \;\;, \label{17}
\end{equation}
with an auxiliary entropy four-current, $\tilde{s}^\mu\equiv su^\mu$, and $\rho _i^\mu\equiv\rho_iu^\mu$

Chemical potentials $\mu_i$ do not imply here traditional chemical equilibrium as they are space and time dependent parameters here. They are related to abelian charges conservation such as the electric charge, the strangeness or the baryon number. In the case of a conserved abelian charge the corresponding chemical potential is written as:
\[\bar{\mu}=\sum\limits_i q_i\mu_i\,,\]%
where $q_i$ denotes the charge of particles of species ''i''.

Taking into account of nonabelian symmetries conservation is more involved and can be formulated on the kinetic equation level \cite{Tur00,Tur02}.

The proper entropy four-current is
\begin{equation}
s^\mu\equiv\tilde{s}^\mu +T^{-1}u_\nu\delta T^{\mu\nu} +\sum_i{\cal L}_i\triangle ^{\mu\nu}\partial _\nu {\cal R}_i \;\;. \label{20}
\end{equation}

\subsection{Second Law of Thermodynamics}
An increasing entropy conditions leads to constraints on the particle source terms $\mathcal{J}_i$ and on the diffusion terms $\mathcal{R}$ and $\mathcal{L}$ \cite{ert}.

The system is driven towards chemical equilibrium if \numparts
\begin{equation}
\sum_i\mu _i{\cal J}_i\leq 0 \;\;.
\end{equation}
This condition is closely related to abelian charge conservation of the system.

The structure of the dissipative term $\delta T^{\mu\nu}$ compatible with $\partial_\mu s^\mu\geq 0$ is given as
\begin{equation}
\delta T^{\mu\nu}=\kappa (\triangle ^{\mu\gamma}u^\nu +\triangle ^{\nu\gamma} u^\mu ){\cal Q}_\gamma +\eta\triangle ^{\mu\gamma}\triangle ^{\nu\delta} {\cal W}_{\gamma \delta} +\zeta\triangle ^{\mu\nu}\partial _\gamma u^\gamma \;\;,
\end{equation}
where $\kappa,\eta$ and $\zeta$ denotes the coefficients of heat conductivity, shear viscosity, and bulk viscosity,  respectively. The heat-flow and the shear tensor  terms  are given here by \eref{heat} and \eref{shear}.

Also, the contribution to the entropy production involving $\mathcal{R}$ becomes a nonnegative if one identifies:
\begin{equation}
{\cal L}_i\equiv -\frac{\mu _i}{T}\;\;,\;\;\;  {\cal R}_i\equiv\sum_j\sigma _{ij}\frac{\mu _j}{T} \;\;,
\end{equation}
\endnumparts
The symmetric matrix $\sigma$ consists of the mutual diffusion constants and has only nonnegative eigenvalues.

The diffusion current
  \begin{equation}
\vec{j}_i\equiv -\nabla {\cal R}_i=-\sum_j\frac{\sigma _{ij}}{T}(\nabla\mu _j-\frac{\mu _j}{T}\nabla T) \;\;.
\end{equation}
involves chemo- and thermo-diffusion contributions.

The entropy production in the relativistic fluid is now given as
\begin{eqnarray}\label{entr_prod}
T\partial_\mu s^\mu &=&-\sum_i\mu_i{\cal J}_i -T\sum_{i,j}\sigma_{ij}(\partial_\mu\frac{\mu_i}{T})\triangle^{\mu\nu} (\partial_\nu\frac{\mu_j}{T}) \nonumber \\ &\;&-\frac{\kappa}{T}{\cal Q}_\mu\triangle^{\mu\nu}{\cal Q}_\nu +\frac{\eta}{2}{\cal W}_\alpha^\beta\triangle_\beta^\gamma {\cal W}_\gamma^\delta\triangle_\delta^\alpha +\zeta (\partial_\mu u^\mu )^2 \,.
\end{eqnarray}
This allows us to specify conditions for particular freeze-outs encountered in high energy heavy ion collisions. The chemical freeze-out is reached with the hadrochemical equilibration. An universal chemical equilibrium means that all sources $\mathcal{J}_i=0$. A partial chemical equilibrium is also possible when only some of sources vanish. That would correspond to the situation when {\it e.g.} light quarks equilibrate faster than heavy quarks. We have then different chemical freeze-outs.

A complete thermal equilibrium is reached when the entropy is not longer produced. It just means $\partial_\mu s^\mu=0$ condition.

\section{Strangeness Production}
We are going to specify terms related to the strangeness production in heavy ion collision processes. Strangeness production in plasma is dominated by gluon fusion $g+g\to s+\bar{s},$ with much smaller contributions of light quarks annihilation processes $q+\bar{q}\to s+\bar{s}.$

We specify the gain and loss equation \eref{gain loss} in the local rest frame
\begin{equation}
\partial _t\rho _i\approx {\cal J}_i\equiv G_s - L_s\,.
\end{equation}
The gain term is given as \numparts
\begin{equation}\label{gain}
    G_s\equiv \langle \sigma_{s}^{gg}v_{gg} \rangle
    (\rho_g)^2
   +\sum_{q=u,d} \langle \sigma_{s}^{q\bar q}v_{q\bar q}
        \rangle\rho_q\rho_{\bar q}\equiv \mathcal{G}\,.
\end{equation}
The loss term is
\begin{equation}\label{loss}
    L_s\equiv [\langle \sigma_{\rm g}^{s\bar{s}}v_{s\bar{s}} \rangle +
\langle \sigma_{q}^{s\bar{s}}v_{s\bar{s}} \rangle]\rho^2_s\equiv\mathcal{L}\rho^2_s\,.
\end{equation}
\endnumparts
The reactions rates $\langle \sigma_{ab} v_{ab}\rangle$ are cross sections averaged over momentum distributions of the components for the gain and the loss process, respectively. The appearance of the $\rho^2_s$ term here instead of the $\rho_s\rho_{\bar{s}}$ term reflects the local strangeness conservation. $\rho_s$ is now the density of produced $s\bar{s}$ pairs.

The source term for strangeness has now a  standard form \cite{KMR} of the kinetic rate equation
\begin{equation}\label{master_bar}
    \frac{\partial\rho_s}{\partial t} =\mathcal{G}-\mathcal{L}\rho^2_s\,.
\end{equation}

Gluons and light quarks numbers are considered to be abundant so we can neglect the change of their numbers due to the strangeness related processes. It means that gluons and light quarks are, in a sense, treated as a heat bath for strange quarks. Similar situation appears also for other heavy quarks processes.

It appears, however, that local exact strangeness conservation leads to strong correlations which should be taken into account \cite{Ko} when \Eref{master_bar} is analyzed. The density $\rho_s$ in this equation is in fact an event-by-event average density. Let's denote event-by-event averages by $\langle\langle\cdot\rangle\rangle$. The rate equation \eref{master_bar} is changed to
\begin{equation}\label{master_evbev}
    \frac{\partial\langle\langle\rho_s\rangle\rangle}{\partial t} =\mathcal{G}-\mathcal{L}\langle\langle\rho^2_s\rangle\rangle = \mathcal{G}-\mathcal{L}\langle\langle\rho_s\rangle\rangle^2 - {\mathcal{L}}\delta\rho^2_s\,,
\end{equation}
where $\delta\rho^2_s$ is the event-by-event fluctuation of the number of $s\bar{s}$ pairs.

The relation of this fluctuation to the event-by-event average of the number of pairs density if crucial for the behavior of the $s\bar{s}$ pairs density. If $\langle\langle\rho_s\rangle\rangle^2 \gg \delta\rho^2_s$ then
 \[\rho_s(\tau)=\sqrt{\frac{\mathcal{G}}{\mathcal{L}}}\tanh(\tau\sqrt{\mathcal{G L}})\,,\]%
-- a  characteristic behavior for the grand canonical distribution.

Increasing fluctuations are related to the exact local strangeness conservation. This effect becomes more profound for strangeness-poor systems. The system would behave - in the simplest scenario \cite{Ko} - according to the canonical distribution. The relation canonical - grand canonical description is more involved in practical applications. It appears \cite{Letessier:2006wn} also that geometric effects should be taken into account as the strangeness distribution is centrality dependent.

\section{Conclusions}
We considered here the entropy producing mechanisms of chemo- and thermo-diffusion together with the contribution of particle production. It was reached within the first order formulation of diffusion process which usually leads to some difficulties. The most serious one is the instability of homogenous equilibrium when even small perturbations of the heat flow, flux velocity or the internal energy density lead system out of equilibrium.  It was pointed out recently \cite{van}  for perfect fluid that by a particular separation of internal and flow energies such instabilities can be removed. It seems also that viscosity and particle production processes added to the standard diffusion processes stabilize a theory and improve the situation even further.

The presented model of strangeness production can be directly extended on other heavy quarks processes. It would be interesting to estimate equilibration time-scales for different flavors at different energies. An implementation of the appropriate numerical code with all initial conditions subtleties is a separate topic. This is necessary to get some quantitative estimations of considered effects. Recent numerical results \cite{Chau06} of the 2+1 shear-viscous one-component hydrodynamics show the quantitative importance of the viscosity in the analysis of the system. Cooling is slower and the transverse expansion is faster. It should be stressed that an inclusion of production processes makes the problem more involved. As was shown \cite{Let94:a} quantitatively, within the thermal model, strangeness production leads to the \emph{faster} cooling of the fluid. This is related to the conversion of the initial condition pressure into strangeness. This effect should be taken account within the hydrodynamical model. The resulting ''effective viscosity'' decreases so the fluid becomes more and more ''effectively perfect``.

An interesting point is the possible dynamical behavior of bulk viscosity terms. The relaxation time for strange and heavy quarks is substantially larger than for light quarks. This increases equilibrium-nonequilibrium ''amplitude``, resulting in the increasing bulk viscosity term.

These problems are objectives of our future research plans.

\ack This work was supported in part by the Polish Ministry of Science and  Higher Education under contract No. N N202 0953 33. The author would like to thank J. Rafelski for inspiring discussions.

\end{document}